\documentclass[aip,graphicx]{revtex4-1}
\usepackage{graphicx}
\draft 

\begin{document}


\title{Anomalous carrier density independent superconductivity in iron pnictides} 



\author{W. Zhou}
\author{X. Z. Xing}
\author{H. J. Zhao}
\author{Z. X. Shi}
\email[]{zxshi@seu.edu.cn;}
\affiliation{Department of Physics and Key Laboratory of MEMS of the Ministry of Education, Southeast University, Nanjing 211189, China}
\author{K. Yamaura}
\email[]{yamaura.kazunari@nims.go.jp}
\affiliation{Superconducting Properties Unit, National Institute for Materials Science, 1-1 Namiki, Tsukuba, Ibaraki 305-0044, Japan}


\date{\today}

\begin{abstract}
Dome-shape superconductivity phase diagram can commonly be observed in cuprate- and iron-based systems via tuning parameters such as charge carrier doping, pressure, bond angle, and etc. We report doping electrons from transition-metal elements (TM = Co, Ni) substitution can induce high-$T_c$ superconductivity around 35 K in Ca$_{0.94}$La$_{0.06}$Fe$_2$As$_2$, which emerges abruptly before the total suppression of the innate spin-density-wave/anti-ferromagnetism (SDW/AFM) state. Unexpectedly, the onset critical temperature for the high-$T_c$ superconductivity stays constant for a wide range of TM doping. Possible extrinsic factors like phase separation, chemical inhomogeneity, and charge carrier cancelation effect are all excluded. This anomalous charge carrier density independent SC is very similar to the interface superconductivity in La$_{2-x}$Sr$_x$CuO$_4$-La$_2$CuO$_4$ bilayer system. The further verified two-dimensional (2D) nature of superconductivity by the Tinkham's angular-dependent critical field model as well as by the angle-resolved magneto-resistance measurements jointly supports the idea of interfacial effect induced high-$T_c$ superconductivity.
\end{abstract}


\maketitle 

The recent observation of high-$T_c$ superconductivity over 100 K in FeSe monolayer film has renewed the interests in Fe-based superconductor (IBSs) studies \cite{ac,ab}. A focus debate ever since its discovery was whether the interfacial effect plays an important role on the giant enhancement of $T_c$ in comparison with FeSe bulk. Interfacial effect enhanced $T_c$ has been commonly reported in cuprate-based heterostructures \cite{t,i,f}. To explain the phenomenon of interface-induced $T_c$ enhancement, many models have been proposed \cite{i,aa,ad,c}. In all models, strong carrier density dependent $T_c$ would be expected \cite{c}. However, this simple principle which both experimentalists and theorists keep following has recently been challenged by the interface superconductivity in La$_{2-x}$Sr$_x$CuO$_4$-La$_2$CuO$_4$ bilayer \cite{i}. It is reported the critical superconducting transition temperature stays essentially constant across a wide doping range, 0.15 $< x <$ 0.47. This surprising results were extracted from an unprecedentedly large set of more than 800 different compositions, which strongly guarantees the experiment repeatability and reliability. This anomalous charge carrier doping independent interface superconductivity poses a big challenge to the ordinary Fermi liquids because a carrier-doping independent chemical potential has to be supposed in its possible explanation.

We report that the similar awkward situation occurs in another system of iron-based compounds, rare-earth-element-doped CaFe$_2$As$_2$. Due to the big $T_c$ difference between (Ca, RE)Fe$_2$As$_2$ (RE = rare earth element) and its counterparts bearing structural and chemical similarities as well as the extremely large magnetic anisotropy, interface effect has previously been proposed for the high-$T_c$ superconductivity origin \cite{d,k,r}. If this assumption is true, (Ca, RE)Fe$_2$As$_2$ system should be the first evidence of interface superconductivity in bulks, which will certainly bring new clues in searching and understanding high-$T_c$ superconductivity. However, further strong evidences for interface superconductivity in (Ca, RE)Fe$_2$As$_2$ is still lack. Through electron doping from TM substitution, we discovered that the primary under-doping Ca$_{0.94}$La$_{0.06}$Fe$_2$As$_2$ without high $T_c$ can be triggered to show superconductivity around $T$ = 35 K. Unpredictably, as the case of interface SC in La$_{2-x}$Sr$_x$CuO$_4$-La$_2$CuO$_4$ bilayer, the high $T_c$ value stays essentially constant against a wide range of TM doping. Benefitted from the easy achievement of TM substitution, a universal and complete electron-doping-dependent high-$T_c$ phase diagram (high-$T_c$ window closed) $T_c$($n$) is established, where $n$ is the net extra doping electron number. The possible extrinsic pitfalls for the strange carrier density independent $T_c$ have been excluded by systematic structural and elemental analysis as well as abundant transport measurements. Simultaneously, through angle- and magnetic field- dependent transport measurement, we give evidence of two-dimensional (2D) SC nature from Tinkham's model of angular-dependent critical field for sufficiently thin films \cite{v}. These experimental data strongly support the existence of interfacial effect in the occurrence of high-$T_c$ superconductivity.

Single crystals of Ca$_{0.94}$La$_{0.06}$(Fe$_{1-x}$TM$_x$)$_2$As$_2$ were grown by FeAs self-flux method as reported before \cite{j,h}. Special attention was paid during the weighting of starting materials to achieve a precise control of the La and TM doping levels. The quality of the obtained crystals has been checked by single crystal X-ray diffraction (XRD) measurement (Supplementary information (SI), S1). Elemental analyses were carried out by energy-dispersive X-ray (EDS) spectroscopy on a field emission scanning electron microscopy (SEM). To accurately determine the doping levels, the compound compositions were determined by average of the multi-point EDS measurements on each crystal. The electrical transport data were obtained by standard four-probe method. For Hall measurements, to cancel the electrode symmetric factor, Hall resistivity is calculated via formula $\rho_{xy}=[\rho_{xy}(\mu_0H > 0)-\rho_{xy}(\mu_0H < 0)]/2$.

\begin{figure}
\includegraphics[width= 8 cm]{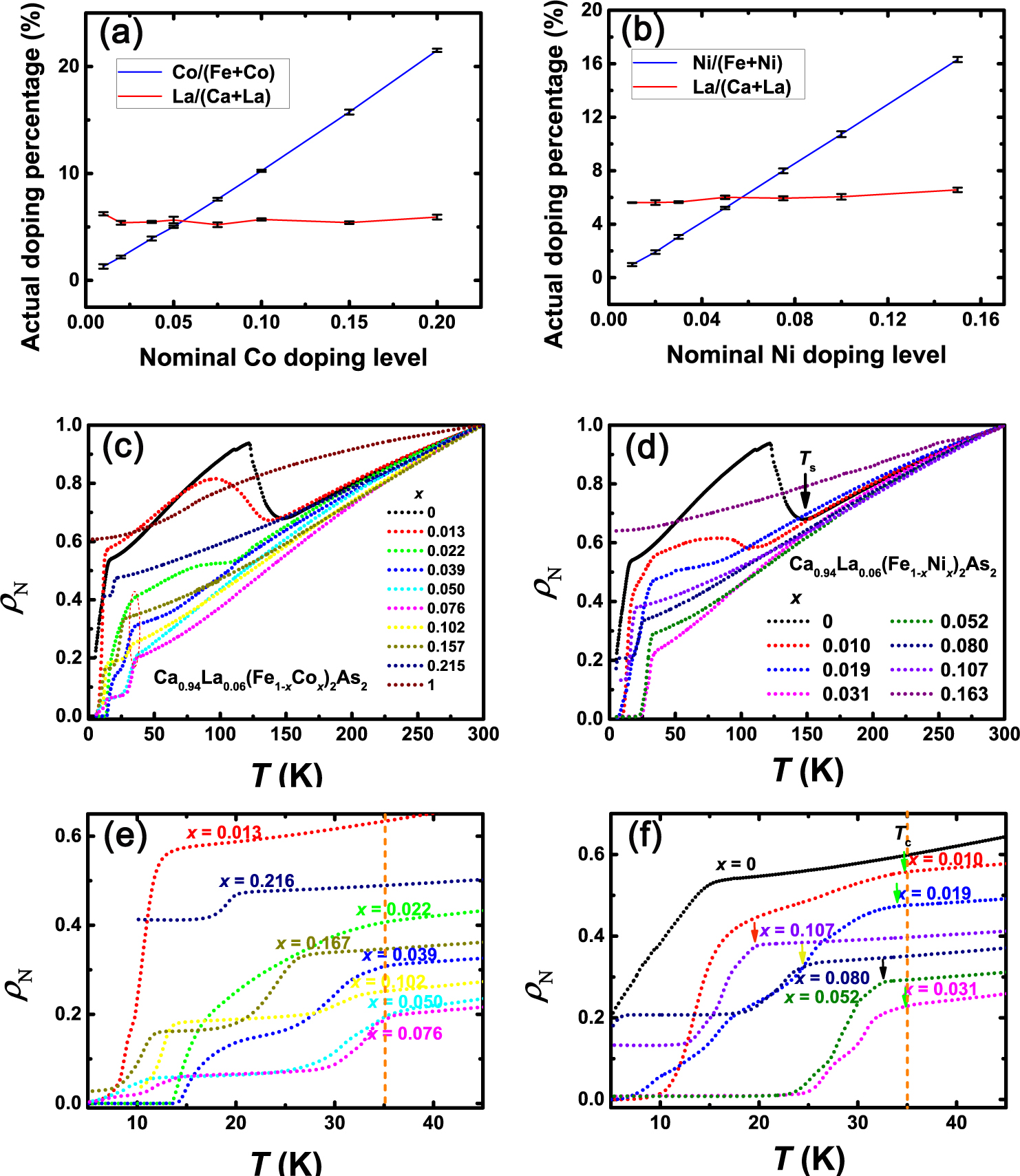}
\caption{\label{} (a-b) The actual doping level of TM (Co, Ni) and La determined by multi-point EDS measurement versus the nominal TM doping level. The error bars are the standard deviations among different measurements. (c-f) Temperature dependence of resistivity ($RT$) curves for Ca$_{0.94}$La$_{0.06}$(Fe$_{1-x}$TM$_x$)$_2$As$_2$ with various $x$. (e) and (f) are the enlarged views of the superconducting transition in $RT$ curves.}
\end{figure}

Figs. 1(a) and (b) show the EDS results. The actual TM doping levels are very close to the nominal values. And as expected, the La doping level keeps invariable as TM doping level $x$ increases. Figs. 1 (c-f) show the corresponding temperature dependence of resistivity ($RT$) for Ca$_{0.94}$La$_{0.06}$(Fe$_{1-x}$TM$_x$)$_2$As$_2$ crystals with different $x$ values, in which the resistivity ($\rho_N$) has been normalized by its value at 300 K. Without TM doping, a SDW transition with a resistivity upturn at $T_s$ $\sim$ 150 K and a low-$T_c$ superconducting transition around 10 K can be identified. As $x$ increases, $T_s$ is gradually suppressed. Simultaneously, SC with high-$T_c$ around 35 K emerges suddenly before the total suppression of SDW state (Diamagnetic signal for 35 K superconductivity is shown in SI, S2). It should be pointed out that the high-$T_c$ SC doesn't directly evolve from the low-$T_c$ phase in low doping levels but emerges abruptly, in stark contrast to the TM doping phase diagrams in many IBSs \cite{n,p}.


\begin{figure*}
\includegraphics[width= 13 cm]{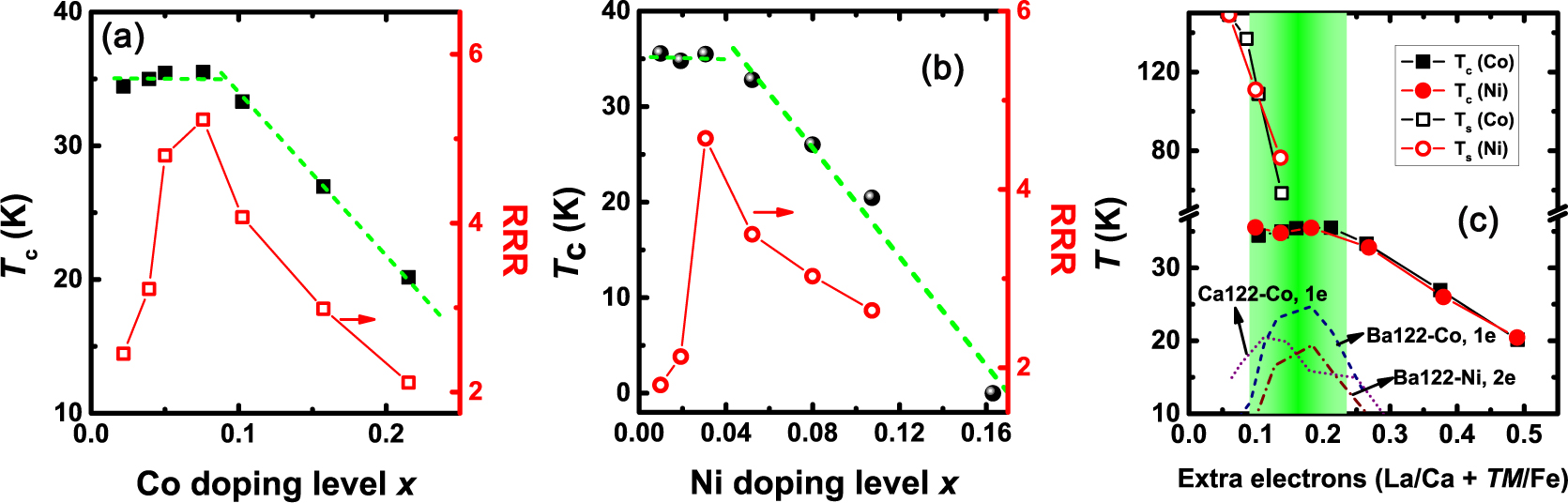}
\caption{\label{} (a-b) Superconducting transition temperature $T_c$ and residual resistance ratio (RRR) plotted as functions of TM doping level $x$. (c) Superconducting and AFM transition temperatures ($T_c$ and $T_s$) plotted as functions of the extra electron numbers ($n$) per unit cell induced by La and TM doping. Every substitution of La for Ca, Co for Fe, and Ni for Fe will introduce 1 e, 1 e, and 2 e, respectively. The dot, dash, and dot-dash lines in (c) represent the dome-shape electron doping dependent superconducting phase diagram for Ca(Fe, Co)$_2$As$_2$, Ba(Fe, Co)$_2$As$_2$, and Ba(Fe, Ni)$_2$As$_2$ in ref. \cite{n,w,y}.}
\end{figure*}

The doping level dependence of the higher superconducting transition $T_c$ is plotted in Figs. 2. Intriguingly, for both the cases of Co and Ni doping, the high-$T_c$ superconducting transition starts at an onset temperature around 35 K and keeps essentially constant for a large TM doping range (see Fig. 2 (a) and (b)). This special range almost covers the superconducting dome in TM-doped BaFe$_2$As$_2$ or CaFe$_2$As$_2$\cite{n,w,y} (see Fig. 2 (c)). In fact, the similar $T_c$-constant behavior has also been noticed in TM-free Ca$_{1-x}$La$_x$Fe$_2$As$_2$ crystals (shown in SI, S3) and in pressure experiments on under-doping Ca$_{1-x}$La$_x$Fe$_2$As$_2$ if the onset $T_c$ is applied \cite{e}. In all cases, after a sufficiently heavy electron doping (or equal effect with electron doping), the onset $T_c$ is found to decrease linearly with $x$. The linear $T_c$($x$) suppression rates are -116.7 K/Fe and -223.8 K/Fe for Co and Ni substitution, respectively. In comparison with Ba$_{0.5}$K$_{0.5}$(Fe, TM)$_2$As$_2$ (suppression rate: -173 K/Fe for Co, -221 K/Fe for Ni), the Ni suppression effect is comparable, while a weaken Co suppression effect is presented \cite{m}. On basis of the rigid-band model, substitution induced electron doping by Co and Ni is expected to be 1 e/Fe and 2 e/Fe, respectively. It is interesting to see that, the $T_c$($n$) curves for Co and Ni doping almost fall into one curve if the net doping electron number from both La and TM doping are taken into account. Both the range of the net doping electron concentration $n$ that high-$T_c$ SC appears and the $T_c$ linear suppression region at heavy TM doping levels are almost identical to each other for Co and Ni substitutions, as shown in Fig. 2 (c). This behavior is in sharp contrast to that in TM-doped BaFe$_2$As$_2$ or CaFe$_2$As$_2$ especially on the following two points \cite{l,n,x}. On one hand, in Ba(Fe$_{1-x}$TM)$_2$As$_2$, since the number of electrons that participate in formation of Fermi surface (FS) is found to decrease from Co to Ni for a fixed nominal extra electron, mismatches among the $T_c$($n$) domes for different kinds of TM doping can be noticed. On the other hand, influenced by the different impurity potentials of the substituted atoms, the electronic structure is possibly changed and thus results in different $T_c$ maximums. Therefore, the $T_c$($x$) constant behavior, the equal maximum $T_c$ values, and the identical linear $T_c$ suppression effect for Co and Ni doping under the rigid-band model are abnormal and unique properties only belonging to the present system.

To understand this unexpected carrier density independent $T_c$, we need to exclude possible experimental artifacts and pitfalls. Specifically, these three scenes, the phase separation, the chemical inhomogeneity, and the charge carrier cancelation effect, come in mind. First, we address the phase separation by systematic XRD and $RT$ measurements. The XRD patterns for TM doping are shown in SI, S1. No crystalline phase separation can be observed. As expected, all (002$l$) peaks gradually move toward higher values with increasing TM substitution for the smaller ion radius of Co and Ni than Fe. The phase separation interpretation should also be untenable in $RT$ curve measurements. Supposing that (Ca, La)Fe$_2$As$_2$ with unknown doping level shows the responsibility for the high $T_c$ and the SDW/AFM transitions in $RT$ curves are only associated to Ca(Fe$_{1-x}$TM$_x$)$_2$As$_2$, the resistivity upturn for SDW transition should persist up to doping level $x = 0.056$ for Co doping and $x = 0.053$ for Ni doping as in ref. \cite{n,x}. However, as shown in Figs. 1 (c) and (d), the disappearances of SDW transitions are much faster in both cases of Co and Ni doping, i.e., the electron concentration contributed by La doping is non-negligible.

To address the second point mentioned above, we emphasize on the elemental analysis. In Figs. 1 (a-b), the standard deviation is shown as the error bar of the actual doping level for every crystal. As can be seen, all error bars are almost too small to be distinguished, which means only small deviations occur among measurements on different points. A table listing raw data of EDS measurement for a typical doping level is provided in the SI, S1, Table I. One can notice the small deviations more intuitively. Given that there is an optimal doping level $x = p$ leading to the maximum $T_c$ as many iron-based systems, we can determine the change strength of the doping-independent range as $(x_{end}-x_{onset})/2p$ where $x_{onset}$ ($x_{end}$) is the onset (end) doping level in the $T_c$-constant region. If we define $p$ as the middle of the doping level in this $T_c$-constant range, the change strength for Co and Ni is calculated to be 55\% and 68\%, respectively, which are far above the standard deviations (usually $<$ 0.2\%) of the element La, the element TM, and even the both. For any element in crystals, no such large or even comparable deviation can be distinguished. Accordingly, we can conclude there is no elemental enriching effect or chemical inhomogeneity that is responsible for the carrier density independent $T_c$.

\begin{figure}
\includegraphics[width= 8.6 cm]{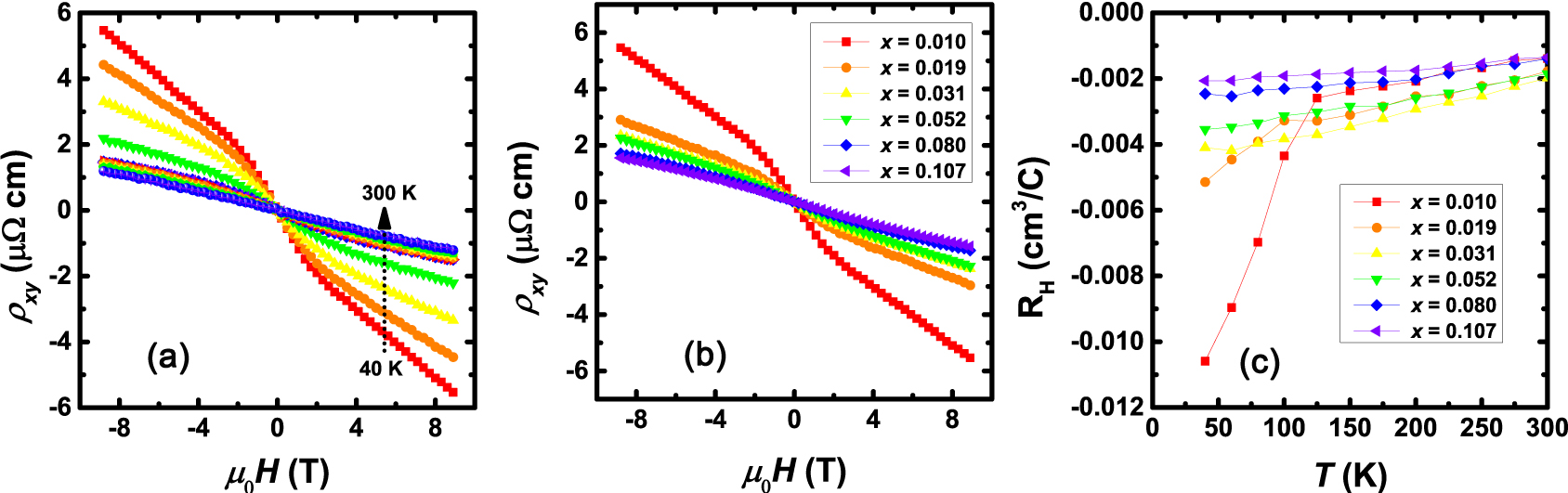}
\caption{\label{} (a) Magnetic field ($\mu_{0}H$) dependence of Hall resistivity ($\rho_{xy}$) for Ca$_{0.94}$La$_{0.06}$(Fe$_{1-x}$Ni$_x$)$_2$As$_2$ with $x$ = 0.010 at various temperatures. (b) Magnetic field dependence of Hall resistivity ($\rho_{xy}$) for Ca$_{0.94}$La$_{0.06}$(Fe$_{1-x}$Ni$_x$)$_2$As$_2$ with various doping levels ($0.010 \leq x \leq 0.107$) at $T$ = 40 K. (c) Temperature dependence of Hall coefficients ($R_{H}$) for Ca$_{0.94}$La$_{0.06}$(Fe$_{1-x}$Ni$_x$)$_2$As$_2$ at low field limits.}
\end{figure}

To address the charge carrier cancelation effect, we performed systematic Hall measurements on Ca$_{0.94}$La$_{0.06}$(Fe$_{1-x}$Ni$_x$)$_2$As$_2$. Fig. 3 (a) shows the magnetic field ($\mu_{0}H$) dependence of Hall resistivity ($\rho_{xy}$) for Ca$_{0.94}$La$_{0.06}$(Fe$_{1-x}$Ni$_x$)$_2$As$_2$ with $x$ = 0.010 at various temperatures. Below the SDW/AFM transition temperature $T_s$, the primary linear $\rho_{xy}$ ($\mu_{0}H$) at high temperatures becomes more and more concave. This behavior may be associated with emergence of mobile charge carriers with high mobility under the SDW transition \cite{b}. The $H$ dependence of $\rho_{xy}$ for Ca$_{0.94}$La$_{0.06}$(Fe$_{1-x}$Ni$_x$)$_2$As$_2$ with different $x$ at $T$ = 40 K are displayed in Fig. 3 (b). The negative slopes indicate the dominant role of electrons in the transport behavior in all crystals. With increasing $x$, the absolute value of the negative slope gradually becomes small. The extracted Hall coefficients ($\text{R}_{\text{H}}$) at low field limit are plotted in Fig. 3 (c). For under-doping crystals ($x$ = 0.010, 0.019), the SDW/AFM transition is also evidenced in $\text{R}_{\text{H}}$($T$) curves. Below $T_s$, $\text{R}_{\text{H}}$ falls down quickly with decreasing $T$, which is a typical signature of the SDW transition observed in many iron-based compounds \cite{w}. Since the high $T_c$ starts from compound $x$ = 0.010 for Ni doping, the existence of SDW state makes a much more drastic change of the electron concentration in this special $T_c$-constant range. This apparently rules out the possibility of the constant net mobile charge carrier (electron) for different $x$ values due to unexpected electronic cancelation effect which may be resulted from As vacancies and other undesired factors \cite{r,g}. Consistent with the $RT$ curves with continuous shape changes, we can conclude the concentration of the mobile charge carrier participating in transport varies drastically and continuously with $x$ increasing.

The exclusion of the above three scenes is also supported by the smooth evolution of the residual resistance ratio (RRR) with doping. It should be noted out that, TM substitution behaves in dual roles, enhancement of impurity density and increasing of electron doping concentration. RRR in TM co-doping samples is a combined result of the scattering density and the mobile carrier concentration. If the transport influence from scattering density enhancement overcomes that from mobile carrier concentration increasing, RRR will increase. Otherwise, RRR will decrease. Due to this opposite influence of the scattering and the mobile carrier concentration on conducting ability, if any of the above three scenes exists, RRR will vary disorderly among different samples and even within samples in a same batch. In this case, no RRR regularity can be obtained. However, as can be seen in Figs. 2 (a) and (b), RRR changes orderly in both cases of Co and Ni substitutions, increasing in the $T_c$-constant region and decreasing monotonously in the linear $T_c(x)$ suppression region. In fact, this evolution of RRR($x$) has been kept in many separated $RT$ curve measurements (also in $RT$ curves for Ca$_{1-x}$La$_x$Fe$_2$As$_2$, see SI, S3.). The initial RRR increase indicates that the effect of mobile carrier concentration increase wins the increase of impurity scattering in influencing transport. While at heavy doping side, the increase of impurity scattering wins the competition. The non-monotonous change of RRR with $x$ increasing indicates a change of the dominated role in influencing transport from mobile carrier concentration to scattering density. Note that, the RRR decrease at heavy doping levels for all cases coincide with the start of linear $T_c(x)$ suppression.

\begin{figure}
\includegraphics[width= 8 cm]{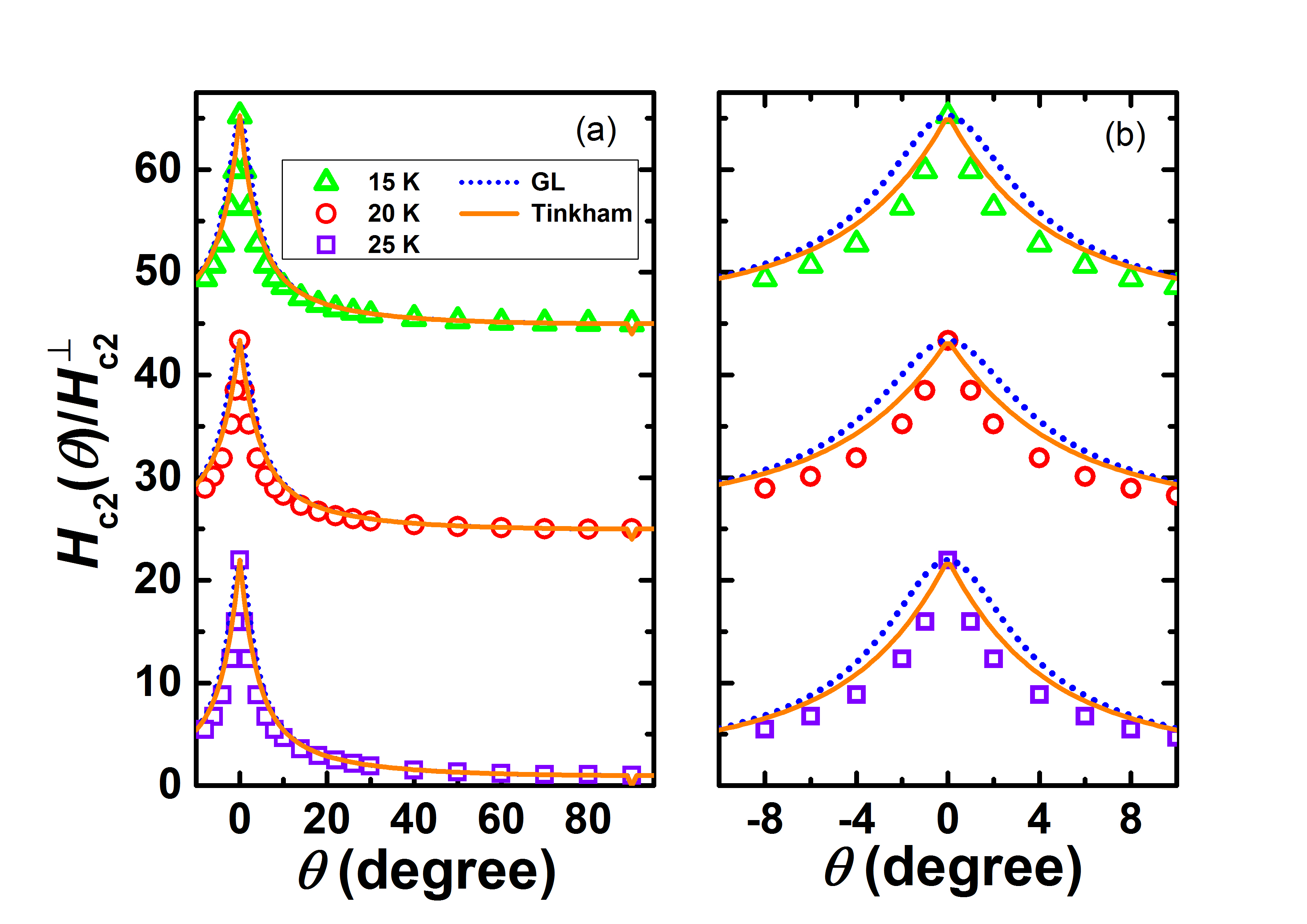}
\caption{\label{} (a-b) The angle $\theta$ dependence of the upper critical field $H_{c2}$ in comparison with $H^{\perp}_{c2}$ for temperatures at 15, 20, and 25 K for a Ca$_{0.94}$La$_{0.06}$(Fe$_{1-x}$Ni$_x$)$_2$As$_2$ ($x$ = 0.052) sample. $\theta$ is angle between the magnetic field orientation and the crystal surface. (b) is an enlarged view for $\theta$ near 0$^{\circ}$. The solid lines represent the reproduced curves based on Tinkham (2D) and GL (3D) formula for angular-dependent $H_{c2}$ using $H^{\perp}_{c2}$ and $H^{\parallel}_{c2}$. For $T$ = 15 K and 20 K (25 K), a criteria $\rho=0.5\rho_{N}$ ($\rho=0.9\rho_{N}$) is used to extract the $H_{c2}(\theta)$ values, where $\rho_{N}$ is the normal state resistivity. The $H_{c2}(\theta)/H^{\perp}_{c2}$ curves for 20 K and 25 K are offset by 24 and 44 along $y$-axis.}
\end{figure}

The unexpected charge carrier density independent $T_c$ is very similar to the situation in La$_{2-x}$Sr$_x$CuO$_4$-La$_2$CuO$_4$ bilayer in which the superconductivity comes from the interface \cite{i}. If the proposed interface SC origin for the high $T_c$ is the right interpretation, one should find the related evidences of 2D SC. As is known, for sufficiently thin superconducting films with thickness $d$ satisfying $d<<\xi_{c}$, Tinkham's formula for angular dependence of the upper critical field $H_{c2}$ is satisfied. Here $\xi_{c}$ is the $c$-axis coherence length. The Tinkham's formula is expressed as \cite{v}
\begin{equation}
\left|\frac{H_{c2}(\theta)\text{sin}\theta}{H^{\perp}_{c2}}\right|+\left(\frac{H_{c2}(\theta)\text{cos}\theta}{H^{\parallel}_{c2}}\right)^2=1.
\end{equation}
$\theta$ is angle between the magnetic field orientation and crystal surface, and $H^{\perp}_{c2}$ ($H^{\parallel}_{c2}$) is upper critical field with field perpendicular (parallel) to the crystal's surface. For 3D bulk superconductors, the angle dependence of $H_{c2}$ can be interpreted by the Ginzburg-Landau (GL) anisotropic mass model with formula $H_{c2}(\theta)=H_{c2}^\parallel (\text{cos}^2 \theta+(H_{c2}^\parallel/H_{c2}^\perp)^2 \text{sin}^2 \theta)^{-0.5}$, which has been frequently applied in IBSs \cite{a}. The most remarkable difference between two models is the slope $|dH_{c2}/d\theta|$ value when $\theta\rightarrow0^{\circ}$. Under Tinkham's model, $|dH_{c2}/d\theta|$ near $\theta\rightarrow0^{\circ}$ is a finite value while it approaches zero in the GL model. We have accordingly measured the resistivity as a function of magnetic field for different field orientations at three selected temperatures below $T_c$ and extracted the angle-dependent $H_{c2}$ using different $H_{c2}$ criteria (see Figs. 4). For field orientation close to crystal's surface, the spaced angle interval has been set closely as 1$^{\circ}$. A cusplike behavior for $\theta$ near 0$^{\circ}$ is observed in $H_{c2}$($\theta$) curves (see Fig. 4 (b)). As expected, 2D Tinkham's model reproduces $H_{c2}(\theta)$ much better than 3D GL model. $|dH_{c2}/d\theta|$ near 0$^{\circ}$ slightly falls below Tinkham interpolation curve which is common to the Nb/Cu samples in the 2D region \cite{u}. In Nb/Cu multilayers, the $H_{c2}(\theta)$ curve near $\theta\rightarrow0^{\circ}$ shows a transformation from a cusplike shape to a rounded shape continuously induced by the thickness ($d_{\text{Cu}}$) change of the Cu layer, from $\xi_{c}=d_{\text{Cu}}$ to $\xi_{c}=6d_{\text{Cu}}$. Therefore, analogically, the superconducting layers in the present system are well separated by non-superconducting region with distance at least comparable to $\xi_{c}$ along $c$-axis. Therefore, the high-$T_c$ superconductivity is two-dimensional in nature. The further evidences verifying the 2D nature of the high-$T_c$ superconductivity from angle-resolved magneto-resistance measurements will be shown in SI, S4.

Recently, two different groups have suggested that the high-$T_c$ superconductivity in (Ca, RE)Fe$_2$As$_2$ may originate crystalline defects from RE dopants \cite{g,r}. Under such scene, the abrupt occurrences of defects and high-$T_c$ superconductivity are still very elusive. In fact, it is argued in ref. \cite{z} that the Pr dopants in the same compounds Ca$_{1-x}$Pr$_x$Fe$_2$As$_2$ do not cluster, but repel each other at short length scales. In the present case, one can naturally speculate that Co and Ni doping are uniform as the case in Ba(Fe, TM)$_2$As$_2$. Then, both Co and Ni doping only induce extra electrons but not promote formation of RE defects. If the idea of defect origin for superconductivity is true, we can conclude that in low-$x$ Ca$_{1-x}$RE$_x$Fe$_2$As$_2$ without high-$T_c$, RE defects already exist. There must has an enough electron concentration for high-$T_c$ occurrence. That is, sufficient electron doping is necessary. Our present data strongly support the conjecture that the unusual enhancement of superconductivity in (Ca, RE)Fe$_2$As$_2$ in comparison with its counterparts may be closely related to some kind of interfacial effect, which is consistent with the recent manifestation of non-negligible interfacial effect on the occurrence of 100 K superconductivity in FeSe monolayer \cite{q}, in which the charge transfer from substrate also contributes to the high-$T_c$. Note that, it is theoretically predicted that quantized superconducting temperature and doping levels could occur in interface superconductivity \cite{c}. Whether there is a correlation between this quantized behavior and the observed $T_c$ constant phenomenon is still an open question.

In summary, we have investigated the TM element substitution effect on the high-$T_c$ superconductivity in (Ca, La)Fe$_2$As$_2$ compound. The onset high $T_c$ has been found to keep almost invariable for a large range of electron doping. Through systematic structural/elemental and transport analyses, the extrinsic scenes, namely the phase separation, the chemical inhomogeneity, and the charge carrier cancelation effect, have been excluded. Additionally, the two-dimensional nature of the high-$T_c$ superconductivity has been verified by magnetic field and angular dependent transport measurements. The present results give strong supports for the interface-effect-induced high-$T_c$ superconductivity in iron pnictides and provide opportunity for exploring new perspective in understanding of especially interface-effect-induced superconductivity.

\begin{acknowledgments}
We are appreciated for the fruitful discussion with Prof. Qianghua Wang in Nanjing university. This work was partly supported by the National Natural Science Foundation of China (Grant No. NSFC-U1432135).
\end{acknowledgments}



%
%

%


\bibliography{TM}

\end{document}